# Quantum algorithm for *de novo* DNA sequence assembly based on quantum walks on graphs.

G. D. Varsamis[1], I. G. Karafyllidis[1,2*], K. M. Gilkes[3], U. Arranz[3], R. Martin-Cuevas[3], G. Calleja[3], J. Wong[3], H. C. Jessen[3], P. Dimitrakis[2], P. Kolovos[4], R. Sandaltzopoulos[4].

1 Department of Electrical and Computer Engineering, Democritus University of Thrace, Xanthi, 67100 Greece
2 National Centre for Scientific Research Demokritos, Athens, 15342 Greece
3. EY Global Innovation Quantum Computing Lab
4. Department of Molecular Biology and Genetics, Democritus University of Thrace, Alexandroupolis, 68100 Greece
*Corresponding author: *ykar@ee.duth.gr*

*Keywords:* DNA sequencing; DNA assembly; Quantum algorithms; Quantum computing

ABSTRACT

*De novo* DNA sequence assembly is based on finding paths in overlap graphs, which is a NP-hard problem. We developed a quantum algorithm for *de novo* assembly based on quantum walks in graphs. The overlap graph is partitioned repeatedly to smaller graphs that form a hierarchical structure. We use quantum walks to find paths in low rank graphs and a quantum algorithm that finds Hamiltonian paths in high hierarchical rank. We tested the partitioning quantum algorithm, as well as the quantum algorithm that finds Hamiltonian paths in high hierarchical rank and confirmed its correct operation using Qiskit. We developed a custom simulation for quantum walks to search for paths in low rank graphs. The approach described in this paper may serve as a basis for the development of efficient quantum algorithms that solve the *de novo* DNA assembly problem.

## 1. Introduction

*De novo* DNA sequence assembly is the assembly of a novel genome for which no reference sequence is available (Imelfort and Edwards, 2009). The aim of the *de novo* sequence assembly is the combination of short reads to form as long as possible continuous sequences (contigs), that encode large parts of the genome under study (Paszkiewicz and Studholme, 2010). In *de novo* assembly, the Reads and their overlaps are used to construct graphs, namely overlap or de Bruijn graphs, where each node represents one Read of the form $R = \{r_1, r_2, \cdots, r_m\}$, with each $r_i$ being a member of the DNA residues type set {A, C, G, T}, and each edge connects those pairs of Reads $(r_i, r_j)$ where $r_j$ can be obtained by shifting the residues of $r_i$ by $k$ places to the left, and adding $k$ new residues at the end of the sequence or, in other words, where a prefix of $r_j$ is equal to a suffix of $r_i$, so that $r_i$ and $r_j$ have an overlap of length *k*. Then, a path that encodes contigs must be found (Rizzi et al., 2019). Thus, the computational aspect of *de novo* assembly is finding Eulerian or Hamiltonian paths, in de Bruijn or overlap graphs, respectively. We considered the case of overlap graphs, where each node represents an error-free and equal-length sequence. This is computationally demanding because finding paths in such graphs is a NP hard problem.

Quantum computers are more effective than their classical counterparts in finding paths in graphs (Dürr et al., 2006; Hastings, 2018). Finding the Hamiltonian path on overlap graph is problem that can be formulated as an Ising Hamiltonian and thus, can be solved using a quantum computer. Since we are in the NISQ era of quantum computation and due to the complexity of the problem, we considered to approach the problem hierarchically and partition it into less complex problems. Hence, we present a novel quantum algorithm with a hierarchical structure for *de novo* DNA sequence assembly. Again, the partitioning of the original graph that describes the problem, can be expressed as an Ising Hamiltonian and thus, can be solved using a quantum computer. The original overlap graph, which we call Graph_1, is partitioned in subgraphs using a maximum cut (MaxCut) quantum algorithm (Fuchs et al., 2021) and a new graph is formed, the Graph_2. The nodes in this graph are sets of nodes that belong to Graph_1. The overlap graphs are expected to have thousands of nodes and the graph Graph_2 is again partitioned in subgraphs that form the graph Graph_3 and so on. In the end, a hierarchical graph structure is obtained.

We use quantum walks to find paths in the graphs of low hierarchical rank and a quantum algorithm to find the Hamiltonian paths in graphs of high hierarchical rank. Quantum walks are a universal quantum computation model equivalent to the quantum gate model (Childs, 2009; Lovett et al., 2010; Childs et al., 2013; Karafyllidis, 2015). Furthermore, quantum walks can also be encoded as quantum circuits and executed on a quantum computer (Douglas, 2009). Several quantum algorithms have been developed for finding Hamiltonian paths in graphs (Hao, 2001, Martoňák et al., 2004). We used the one which is also used in Qiskit (Srinivasan, 2018). We tested the correct operation of our algorithm by simulating it using Qiskit. The quantum algorithm can serve as a basis for efficient and fast *de novo* DNA sequence assembly.

## 2. Outline of the quantum algorithm.

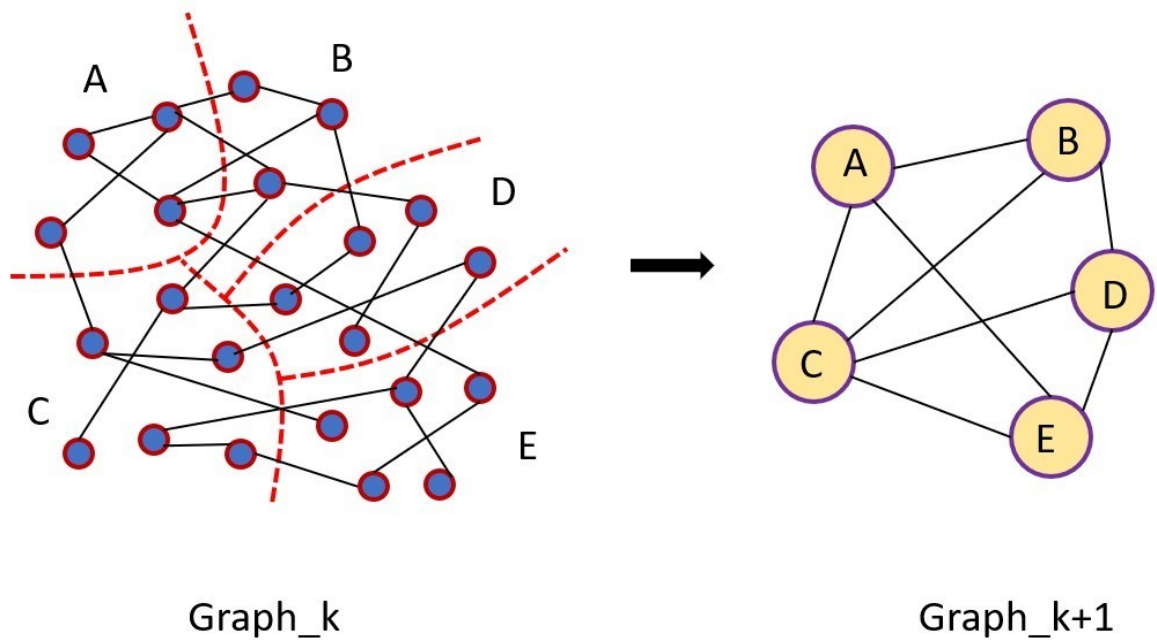

*Figure 1. Schematic description of the overlap graph hierarchical structure. The graph with rank Graph_k is partitioned in five parts: A, B, C, D and E, which form the nodes of the graph Graph_k+1.*

The Read overlap graph is the basic graph and its rank is Graph_1. Graph_1 is partitioned in subgraphs using a MaxCut quantum algorithm. Each partition will be a node of the graph Graph_2 and the edges between two partitions of Graph_1 will form an edge between the corresponding nodes of Graph_2. The weight of this edge is equal to the sum of the weights of the edges connecting the two partitions in Graph_1. Figure 1 shows a schematic description of the overlap graph hierarchical structure. In this figure, the graph with rank Graph_k is partitioned in five partitions labeled A, B, C, D and E. These partitions are the nodes of the graph Graph_k+1. We use quantum walks to find contigs in the low rank graphs, which contain large numbers of nodes and find the contigs in the high rank graphs using a quantum algorithm for Hamiltonian paths. So, instead of finding the maximum contig from the overlap graph, which is a very difficult task, we find a contig in each graph in the hierarchy. Combining these graphs will produce a set of contigs, one of which will be the maximum contig that assembles the full-length sequence in question.

Our quantum algorithm consists of three subroutines. The first one is the MaxCut quantum algorithm which is available in Qiskit. The second is a quantum algorithm that finds Hamiltonian paths, which is also available in Qiskit. The third subroutine is a quantum walk algorithm which we developed that finds contigs in graphs. In this algorithm, the weights of the edges are expressed as potential differences between the nodes.

**3. Quantum walks in overlap graphs.**

Quantum walks are quantizations of classical random walks. The quantum walk is not random but can by directed according to the parameters of the specific problem. A particle, the quantum walker moves in a graph, the nodes of which are the basis states of a quantum register. If the graph has $N$ nodes, a quantum register with $n$ qubits is needed to label the nodes with its basis states, where $N = 2^n$. This register is the position quantum register. The size of the graphs in the hierarchy is chosen according to the qubits that are available. Our quantum algorithm is scalable because the graph size can increase as the number of available qubits increases. The basis states of the quantum register, $|n\rangle$, are expressed in the computational basis which spans a Hilbert space, $H_B$. The direction of motion in the graph is determined by the basis states of a second quantum register, $|m\rangle$, called the coin quantum register. If the maximum number of edges connected to nodes is $M$, then this quantum register contains $m$ qubits, where $M = 2^m$. The basis states of this register span a small Hilbert space $H_D$ (Varsamis et al., 2022, Varsamis and Karafyllidis, 2023). The quantum walk evolves in a Hilbert space, $H$, which is the tensor product of the position and coin Hilbert spaces:

$$H = H_B \otimes H_D \qquad (1)$$

The quantum walk is controlled by two unitary operators, the coin and motion operators. These operators can be constructed using quantum gates (Karafyllidis, 2015). The coin operator, $\hat{C}$, acts on the coin register basis states and is represented by an $mXm$ unitary matrix:

$$\hat{C} = \begin{bmatrix} c_{11} & c_{12} & \cdots & c_{1m} \\ c_{21} & c_{22} & \cdots & c_{2m} \\ \vdots & \vdots & \vdots & \vdots \\ c_{m1} & c_{m2} & \cdots & c_{mm} \end{bmatrix} \qquad (2)$$

The action of $\hat{C}$ on the coin basis states of the coin register $|m\rangle$ is given by:

$$\hat{C}|m\rangle = \sum_{x=1}^{m} \sum_{i=1}^{m} c_{i,x} |x\rangle \qquad (3)$$

The motion operator $\widehat{M}$ acts on the basis states of the position register, $|n\rangle$, and moves the quantum walker from node to node through the graph. We consider the general case where the quantum walker is located on the node $|n\rangle$ of the overlap graph. There are $m$ edges connected to this node. The $\widehat{M}$ operator moves the quantum walker to a neighboring node $|p\rangle$ according to the probability amplitude of the coin register state $|q\rangle$ that corresponds to the motion via the edge connecting the two nodes:

$$\widehat{M} |n\rangle = \left(\sum_{i=1}^{m} c_{i,q}\right) |p\rangle \qquad (4)$$

In overlap graphs, the nodes are connected by weighted edges. The weight of each edge is determined by the number of common residues in the Reads that are encoded as nodes. We want the quantum walk to evolve along edges with maximum weight for the path to encode the contig. To obtain this, we introduce potential differences between the nodes of the overlap graph, which are equal to the weight of the edges. By introducing these potential differences, the $\widehat{M}$ operator will move the quantum walker via edges with the maximum number of common residues.

Quantum walks are described by the Schrödinger equation, in which the Hamiltonian operator is the sum of the kinetic and the potential energy:

$$\widehat{H} = -\frac{\hbar^2}{2m} \frac{\partial^2}{\partial x^2} + V(x,t) \qquad (5)$$

A special solution to the Schrödinger equation is:

$$\widehat{U} = exp\left(\frac{it}{\hbar} \frac{\hbar^2}{2m} \frac{\partial^2}{\partial x^2}\right) exp\left(\frac{-it}{\hbar} V(x,t)\right) \qquad (6)$$

The first exponential of the unitary operator $\widehat{U}$ corresponds to the kinetic energy part, which in the quantum walk model is expressed by the move operator $\widehat{M}$. The second exponential corresponds to the potential energy. If $R$ is the number of common residues between two nodes $|n\rangle$ and $|p\rangle$ is $V(n,p) = R$. In our algorithm, when the quantum walker moves from node $|n\rangle$ to node $|p\rangle$ peaks up a phase factor that depends on the value of the potential $V(n,p)$:

$$\exp(i\,\varphi(n,p)) = exp\left(\frac{-it}{\hbar} V(n,p)\right) \qquad (7)$$

This potential dependent phase factor is introduced using phase quantum gates. Using the formulation described above, the quantum walker moves in the overlap graph along edges with maximum weight thus tracing the path that corresponds to the contig.

## 4. Simulation of the quantum algorithm

To simulate the quantum algorithm described above, we considered an artificial dataset, consisting of eight Reads. First, we constructed the overlap graph (Graph_1) as presented in Fig. 2. The red nodes represent each Read and the edge weights correspond to the overlap, as mentioned above.

| Reads | Node |
|---|---|
| AAGCTTCCA | 0 |
| TTCCAGTGC | 1 |
| TGCAGGCAT | 2 |
| GCATCCGGA | 3 |
| CGGACATGG | 4 |
| CATGGCTAA | 5 |
| TAAGCAATT | 6 |
| AATTCCGTA | 7 |

(a)

| Connected nodes | Overlap |
|---|---|
| 0 - 1 | 5.0 |
| 1 - 2 | 3.0 |
| 1 - 3 | 2.0 |
| 2 - 3 | 4.0 |
| 3 - 4 | 4.0 |
| 4 - 5 | 5.0 |
| 5 - 6 | 3.0 |
| 5 - 7 | 2.0 |
| 6 - 7 | 4.0 |

(b)

*Table 1. (a) Artificial Read dataset and their corresponding nodes. (b) Overlap between reads*

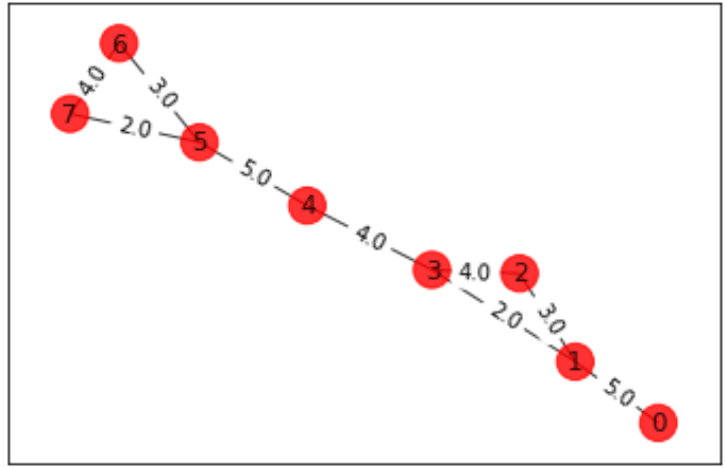

*Figure 2. Overlap graph. Node 0: **AAGCTTCCA**, node 1: **TTCCAGTGC**, node 2: **TGCAGGCAT**, node 3: **GCATCCGGA**, node 4: **CGGACATGG**, node 5: **CATGGCTAA**, node 6: **TAAGCAATT**, node 7: **AATTCCGTA**.*

The next step of the algorithm is to partition the overlap graph (Graph_1). To do so, we created the adjacent matrix of the graph, where we used $1/overlap$ for connected nodes and a value $\geq 1$ for non-connected nodes. The corresponding adjacency matrix is:

$$A = \begin{bmatrix} 1.0 & 1/5 & 1.0 & 1.0 & 1.0 & 1.0 & 1.0 & 1.0 \\ 1/5 & 1.0 & 1/3 & 1/2 & 1.0 & 1.0 & 1.0 & 1.0 \\ 1.0 & 1/3 & 1.0 & 1/4 & 1.0 & 1.0 & 1.0 & 1.0 \\ 1.0 & 1/2 & 1/4 & 1.0 & 1/4 & 1.0 & 1.0 & 1.0 \\ 1.0 & 1.0 & 1.0 & 1/4 & 1.0 & 1/5 & 1.0 & 1.0 \\ 1.0 & 1.0 & 1.0 & 1.0 & 1/5 & 1.0 & 1/3 & 1/2 \\ 1.0 & 1.0 & 1.0 & 1.0 & 1.0 & 1/3 & 1.0 & 1/4 \\ 1.0 & 1.0 & 1.0 & 1.0 & 1.0 & 1.0 & 1/4 & 1.0 \end{bmatrix} \quad (8)$$

We utilized Qiskit's MaxCut module to partition Graph_1 using the Quantum Approximate Optimization Algorithm (QAOA) (Farhi et al., 2014). The circuit of QAOA and the partitioned graph are presented on Fig. 3 and Fig. 4 accordingly. After the partitioning, Graph_2 is created where each one of its nodes contains the corresponding nodes of Graph_1. On Fig. 4(d) node 0 corresponds to node 4 on Graph_1, node 1 to node 5 and so on.

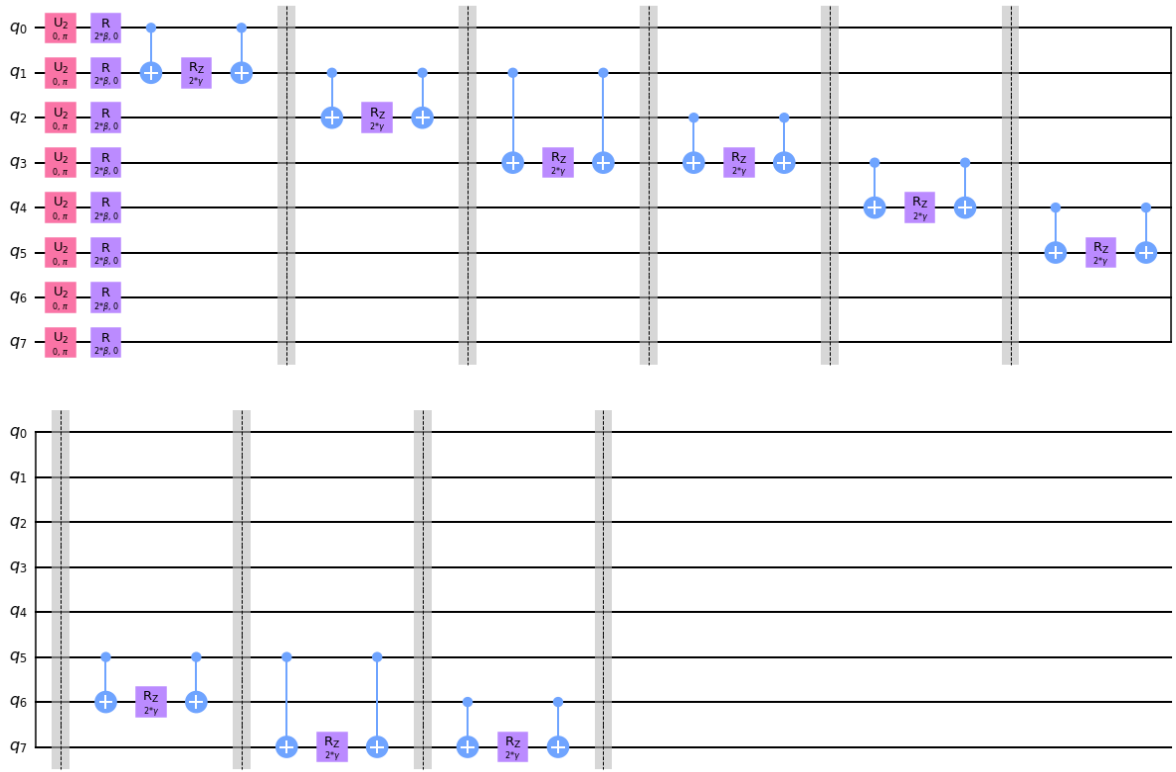

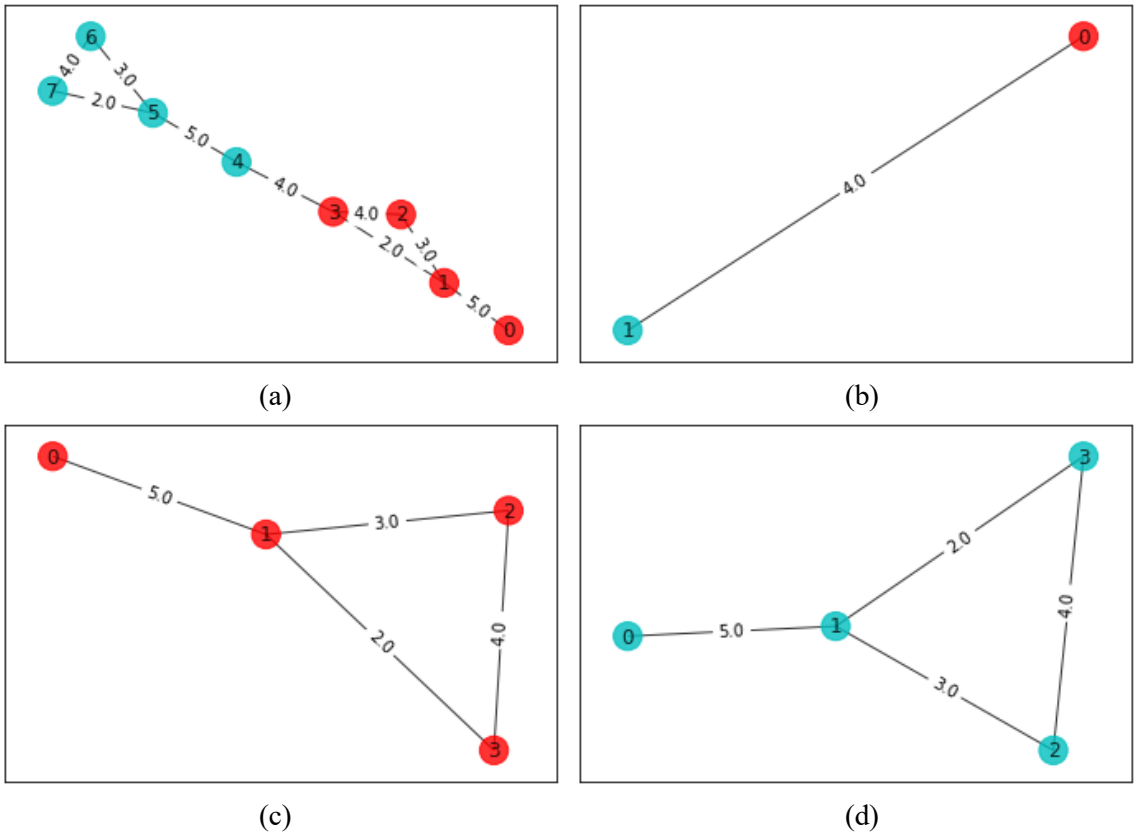

Figure 3. QAOA circuit for the partitioning of Graph_1

(a)

(b)

(c)

(d)

Figure 4. (a) Partitioned graph. (b) Graph_2. (c) Nodes contained on Graph_2's node 0 (red). (d) Nodes contained on Graph_2's node 1 (cyan).

Considering the graph presented on Fig. 4(c), we utilized discrete time quantum walks evolving on potential spaces to find the Hamiltonian path. We extracted the corresponding adjacent matrix following the steps that produced Eq. (8) and we mapped it as the applied potential, on a 4x4 lattice that serves as the evolution space for the quantum walker (Fig. 5(a)). We utilized the main diagonal of the lattice as the benchmark for our calculations. Thus, we set the quantum walker on an equal superposition on the main diagonal lattice sites, let it evolve and finally calculated the probability of finding the walker on each site on the main diagonal. The highest probability corresponds to the first node of the path and the lowest probability to the last (Fig. 5(b)). Lattice site (0,0) corresponds to node 0, (1,1) to node 1 and so on. Our aim is to maximize the overlap that comes from the predicted path. The Hamiltonian path that maximizes the overlap on the graph of Fig. 4(c) is presented on Fig. 5(c).

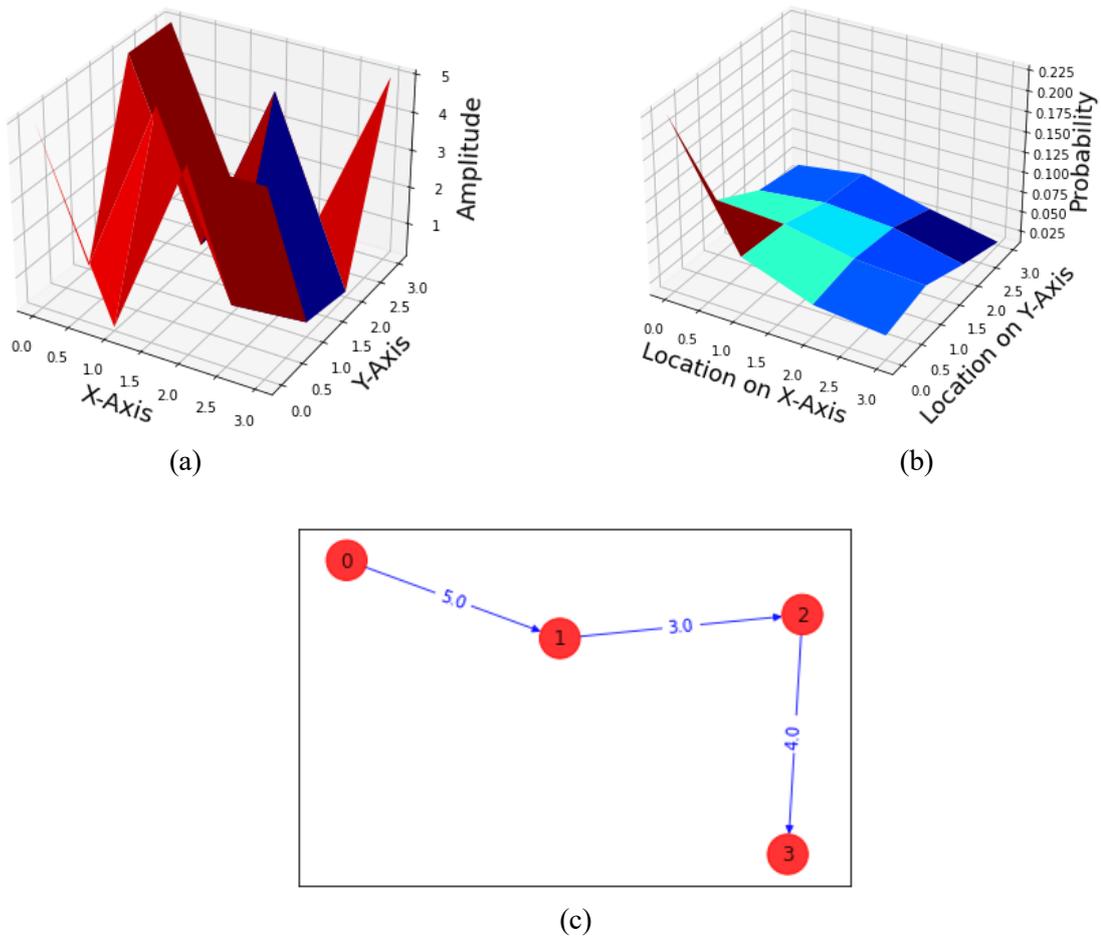

Figure 5. (a) Adjacency matrix as applied potential on the evolution lattice. (b) Probability distribution of the quantum walk evolution (c) Maximum overlap Hamiltonian path.

Due to the symmetric nature of the graph presented on Fig. 4(d), the corresponding Hamiltonian path is the same as the one presented on Fig. 5(c). The same results can be reproduced using QAOA on Qiskit. The optimization problem of finding the Hamiltonian path, corresponds to a Traveling Salesman Problem (TSP) and can be described as a quadratic problem with the corresponding constraints (Lucas, 2014). Mapped as an Ising Hamiltonian, it can be solved on a quantum computer. Yet, as the number of nodes increases, so does the complexity and, hence, quantum walks produce better results.

Finally, after finding the Hamiltonian paths for the graphs contained on the nodes of Graph_2 (Fig. 4(b)) we used QAOA (Fig. 6(a)) to calculate the Hamiltonian path for Graph_2. (Fig. 6(b))

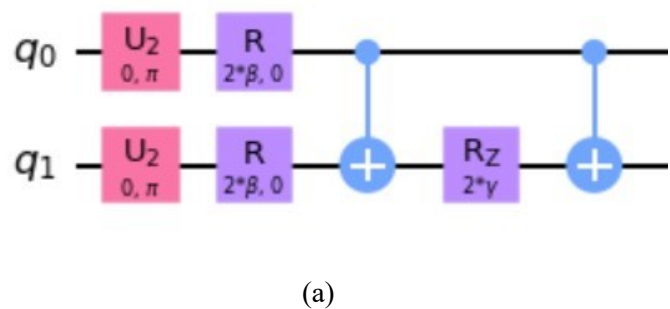

(a)

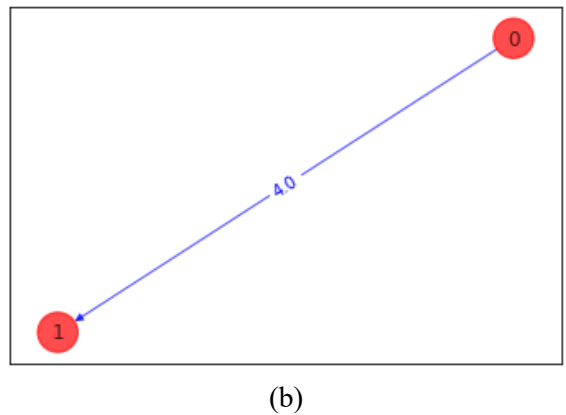

(b)

*Figure 6. (a) QAOA circuit for Graph_2. (b) Hamiltonian path for Graph_2*

Tracing back the corresponding Hamiltonian paths found in the graphs contained in Graph_2's nodes, the Hamiltonian path of the original overlap graph, Graph_1, is 0 → 1 → 2 → 3 → 4 → 5 → 6 → 7 (Fig. 7). Taking into account that these nodes correspond to genome Reads, the final assembled genome would be:

AAGCTTCCAGTGCAGGCATCCGGACATGGCTAAGCAATTCCGTA

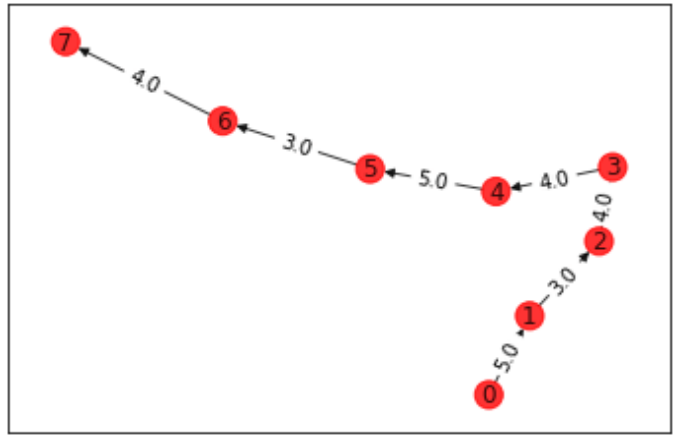

*Figure 7. Maximum overlap Hamiltonian path for Graph_1. Corresponding assembled genome AAGCTTCCAGTGCAGGCATCCGGACATGGCTAAGCAATTCCGTA*

**5. Conclusions**

We presented a novel quantum algorithm for *de novo* DNA sequence assembly. We introduced the partition of the overlap graph and formed a hierarchical structure using the MaxCut quantum algorithm. We developed a quantum algorithm based on quantum walks in graphs, where the edge weights are modeled as potential differences, and used it to find contigs in low rank graphs. We used a quantum algorithm, available in Qiskit, to find the contigs in high rank graphs expressed as Hamiltonian paths. We verified the correct operation of our algorithm by simulating it in Qiskit. The quantum algorithm is scalable and can be extended to assemble larger sequences as the number of available qubits increases.

**Acknowledgments**